\documentclass[aps,prb,twocolumn,longbibliography]{revtex4-1} 
\usepackage{amsmath} 
\usepackage{graphicx} 
\usepackage{SIunits} 

\begin{document}

\title{Motion Tracking in Undergraduate Physics Laboratories with the Wii Remote} 
\author{Spencer L. Tomarken} \email{slt@uchicago.edu} \affiliation{Department of Physics, The University of Chicago, Chicago, IL 60637} 
\author{Dallas R. Simons} \affiliation{Department of Physics, Harvard University, Cambridge, MA 02138} 
\author{Richard W. Helms} 
\author{Will E. Johns} 
\author{Kenneth E. Schriver} 
\author{Medford S. Webster} \affiliation{Department of Physics and Astronomy, Vanderbilt University, Nashville, TN 37235}

\begin{abstract}

We report the incorporation of the Wiimote, a light-tracking remote control device, into two undergraduate-level experiments. We provide an overview of the Wiimote's basic functions and a systematic analysis of its motion tracking capabilities. We describe the Wiimote's use in measuring conservation of linear and angular momentum on an air table, and measuring the gravitational constant with the classic Cavendish torsion pendulum. Our results show that Wiimote is a simple and affordable way to streamline the data acquisition process and produce results that are generally superior to those obtained with conventional techniques. 
\end{abstract}

\maketitle

\section{Introduction}

Since its introduction in 2006, Nintendo's \emph{Wii} game console has garnered considerable attention for its novel controller, the Wii Remote (Wiimote).\cite{nintendo} With solid-state accelerometers and a front-mounted camera, the Wiimote incorporates the controller's motion and position into the user interface. For example, a player can swing the Wiimote like a baseball bat or use it as a pointer for selecting on-screen options.

The Wiimote communicates with the console over a standard wireless Bluetooth interface, making it quite simple to use the remote with any Bluetooth-enabled PC. The range is typically shorter than the range of Wi-Fi, but adequate for classroom-scale projects, and data transfer is virtually instantaneous for the kinds of applications we describe in this paper.

It is not surprising, therefore, that the Wiimote has found applications in instructional physics laboratory experiments, with most of the attention focused on using its accelerometers.\cite{Vannoni2007,ochoa:16,kawam:508} In this paper we describe how to use the Wiimote for tracking the motion of objects in two common undergraduate laboratory experiments. The first experiment demonstrates conservation of linear and angular momentum using elastic collisions on an air table. The second experiment updates the familiar Cavendish measurement of the gravitational constant.

In both cases, Wiimote data acquisition offers a fun alternative for students already fond of the Wii game system. Moreover, the data recorded is typically much greater, in quantity and quality, than what can be obtained with more traditional methods. This affords students the opportunity to follow up their experiment with more sophisticated off-line data-analysis techniques.

\section{Wiimote Basics}

When playing a typical Wii game, the Wiimote's accelerometers provide the console with partial information about the controller's spatial orientation. Additional information, such as whether a Wiimote is pointed at the right or left side of the screen, requires information from its camera. The user places an array of IR LEDs, the \emph{sensor bar}, next to the television or monitor and points the Wiimote at a sequence of on-screen targets. To avoid being distracted by the screen or other light sources, the Wiimote's camera sits behind an IR filter. The only thing the Wiimote \emph{sees} is the sensor bar.

The power of the Wiimote is that it does the image processing \emph{before} sending the data to the console. It returns the \emph{coordinates} of the sensor bar LEDs, not the actual camera image. The console software then builds a map, connecting the screen coordinates of the targets with the reported camera coordinates of the LEDs. When the mapping is complete, the console can determine where the user is pointing to, as long as the sensor bar remains within the Wiimote's field of view.

Our experiments reverse the roles of the Wiimote and the sensor bar. Rather than using fixed lights as a reference system for a moving Wiimote, we fix the Wiimote and use it to track moving lights. Tracking invisible IR LEDs is inconvenient, but the IR filter is easily removed by opening the Wiimote housing with a tri-wing screwdriver. Even with normal overhead lighting on, we have found that the Wiimote can track LEDs of all colors, as well as spots from red and green laser pointers, dots on a computer monitor, and even reflective stickers.

\begin{figure}
	\includegraphics{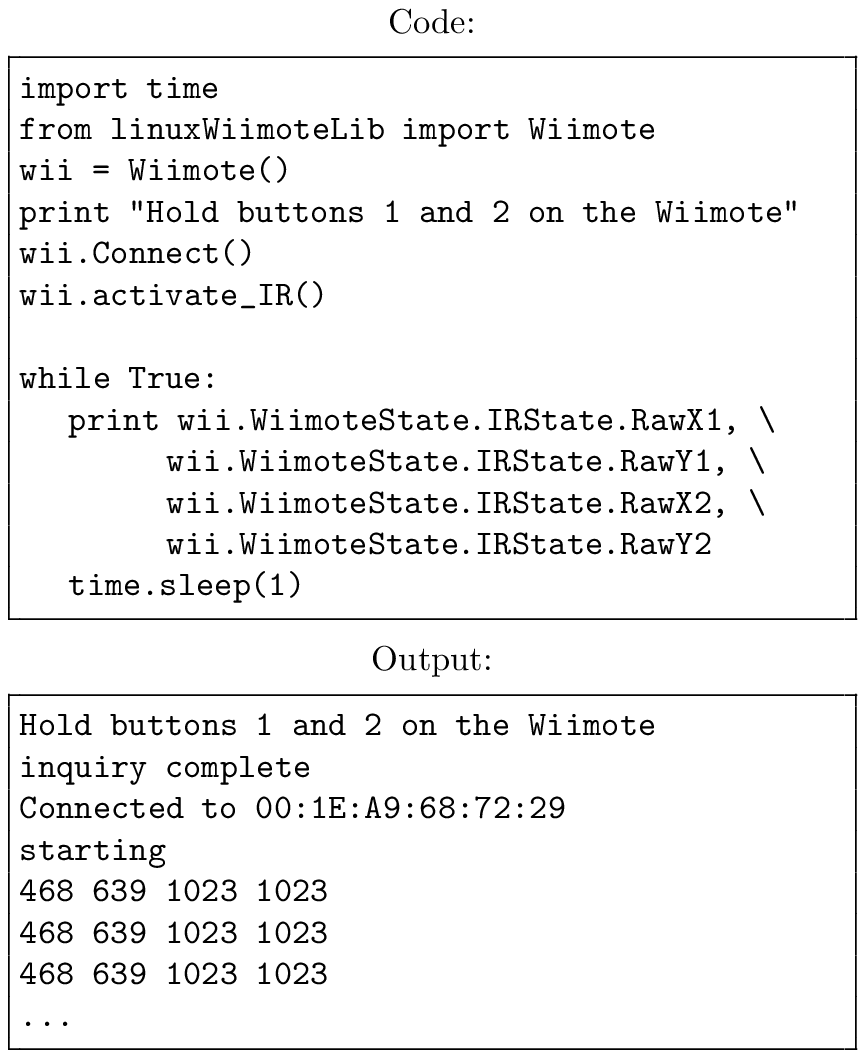}
\caption{Basic Python program for extracting Wiimote coordinates. This program reports the $x$ and $y$ coordinates of the first two spots being tracked by the Wiimote at one-second intervals. A value of 1023 indicates no spot is being tracked by that slot.} 
\label{fig:samplecode} 
\end{figure}

Bypassing the console and extracting the light coordinates directly from the Wiimote is surprisingly simple. Doing this is so popular that numerous software libraries are freely available, allowing any Bluetooth-enabled computer to talk to a Wiimote. Our work utilized a free, slightly modified library, written in Python for Windows and Linux.\cite{gtk2008} As shown in Fig.~\ref{fig:samplecode}, once the right supporting libraries are in place, a few lines of code are sufficient to track lights and collect the data on a PC.

The Wiimote returns the $x$ and $y$ coordinates of up to four tracked lights with a resolution of $1024\times768$~pixels. Its horizontal field of view is approximately $\pm25\degree$. However, the Wiimote has no way to distinguish one light source from another. It acquires them in an essentially random order and assigns them to four slots. If a source is lost and reacquired, it will be assigned to the lowest available slot, not necessarily to the one it previously occupied. This behavior is rare in our experience unless a stray hand blocks a light.  Various schemes can fix this problem during data acquisition, if so desired.

\section{Testing Motion Capture Performance}

Since the demands of an experiment may exceed those of a video game pointer, we have conducted several tests of the Wiimote motion tracking performance. To test whether the Wiimote's image capture apparatus introduces any nonlinear distortion of the spot position, a Wiimote was mounted on an optics bench in front of an LCD monitor. The distance between the Wiimote and the screen was such that the edges of the monitor barely exceeded the Wiimote's field of view, and the screen and Wiimote were leveled and squared.

\begin{figure}
	\includegraphics{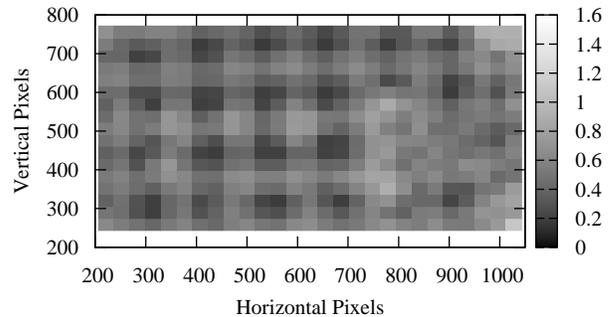} 
\caption{Deviation from linearity over Wiimote's field of view. The Wiimote was placed in front of an LCD screen showing a precisely-controlled moving spot. The spot was scanned across the Wiimote's entire field of view, and the screen coordinates were mapped linearly onto the Wiimote's coordinates. The graph shows the fit residual $\left(\sqrt{\delta_x^2 + \delta_y^2}\right)$.} 
\label{fig:linearity} 
\end{figure}

A single white dot, with a width of about 2\% of the Wiimote's field of view, was scanned across the LCD monitor, and the known screen coordinates at each position were compared with the coordinates returned by the Wiimote. In order to account for differences in scale between Wiimote pixels and screen pixels, a linear fit was performed between the two sets of coordinates in both horizontal and vertical directions according to
\begin{align}
	x_\text{Wii} &= a_1 + b_1 x_\text{scr} + c_1 y_\text{scr} \\
	y_\text{Wii} &= a_2 + b_2 x_\text{scr} + c_2 y_\text{scr}.
\end{align}
The residuals from the two fits are combined in quadrature and shown in Fig.~\ref{fig:linearity}. The RMS residuals for the horizontal and the vertical fits were both 0.39 pixels. Given that the expected error due to integer truncation is $0.29$ pixels, this result is quite satisfactory. However, an interesting feature in Fig.~\ref{fig:linearity} is the periodic beating due to roundoff between the two different pixel sizes.

The next test determined how the Wiimote assigns coordinates to a spot that extends over several camera pixels. Using the same setup as the previous experiment, the Wiimote was shown a stationary dot of increasing radius. The reported coordinates were identical for dots as large as 10\% of the Wiimote's field of view, indicating that the Wiimote tracks the center, rather than the perimeter, of large spots. For extremely small dots, brightness is likely the limiting factor. 

\begin{figure}
	\includegraphics{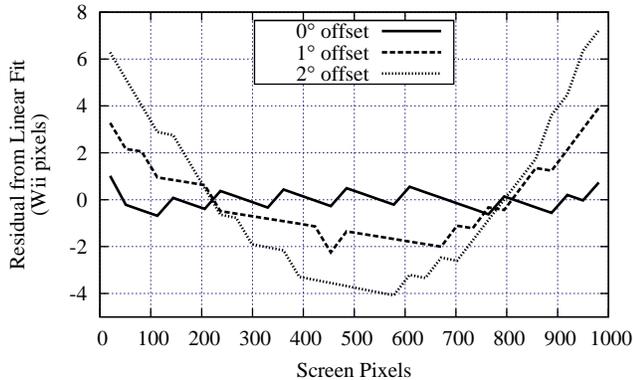} 
\caption{Deviations from linearity introduced by Wiimote yaw. Using the same setup as Fig.~\ref{fig:linearity}, the spot was scanned once horizontally across the the Wiimote's field of view. The Wiimote was given yaw rotations of $-1\degree$, $0\degree$, $+1~\degree$. The graph shows the fit residual $\delta_x$} 
\label{fig:misalignment}
\end{figure}

Finally, the tolerance for misaiming the Wiimote was characterized by rescanning the dot with the Wiimote rotated horizontally (yaw). As Fig.~\ref{fig:misalignment} shows, this kind of misalignment produces the expected quadratic nonlinearity, also called keystone distortion, between screen coordinates and Wiimote coordinates. These distortions must be considered during experimental design. Using the Wiimote's full field of view improves resolution but amplifies the effects of misalignment-induced nonlinearities.

\section{Tracking Objects -- Conservation of Momentum} 

In our first example of Wiimote data acquisition, students observe conservation of linear and/or angular momentum with colliding objects on a low-friction air table. The data for this popular experiment is typically recorded by a video camera.\cite{williamson:2000} Students use software to tag the moving objects in individual frames and us this information to calculate the velocity of each object.

The Wiimote can simplify this process by tracking the moving objects directly, removing the need for inspecting individual video frames. To make the objects \emph{trackable}, their tops are covered in black paper with a reflective sticker attached to the center. If one also wishes to track the rotation of an object, then an additional sticker can be attached at its edge. Attaching two stickers to one object, separated by a known distance, also provides a calibration to convert Wii pixels into physical lengths. The Wiimote is mounted above the air table such that as much of the table as possible is within the Wiimote's field of view.

Reflective stickers are convenient because they require no power source and add negligible mass to the colliding objects. When used with only standard fluorescent lighting, the stickers' reflective brightness appears to be at the limit of what the Wiimote can distinguish from ambient light sources. In order for this experiment to work reliably, the Wiimote's gain was increased through a modification of our software library (see App.~\ref{sec:app}).

In our particular version of this experiment, students collide a circular puck with a rectangular block. Although the block starts at rest, it has both rotational and translational motion after the collision. This shows that, for a suitably chosen reference point, an object traveling in a straight line possesses angular momentum.

Figure 4 shows the data collected for one such collision, along with the reconstructed positions of the colliding objects. The computer polls the Wiimote at a rate chosen by the user (here 50 Hz) and records the returned coordinates and the computer's elapsed time. The rotation of the puck was negligible and was not tracked or included in the calculations. In principle, it could be included  using an extra spot, just as with the block. Table~\ref{tab:momentum} shows student data for a particular run, and shows that both linear and angular momentum are conserved to better than 2\%. Since the Wiimote alignment relative to the air table was very crude, improved accuracy is quite possible.

The initial velocity of the puck was approximately 30~\centi\meter\per\second, which for our setup corresponds to 400~pixels/s. Although we did not systematically test the Wiimote's ability to track faster objects, our experience suggests that it should be possible.

\begin{table}
	\caption{Results of studies of conservation of linear and angular momentum for a collision between two objects on an air table.  The positions of the objected were recorded by the Wiimote.} 
	\begin{ruledtabular}
		\begin{tabular}
			{lrrrr} & Puck & Block & Total & \\
			\hline $p_{x,\text{before}}~(10^{-3}~
			\newton\,\second)$ & -16.862 & -0.062 & -16.924 & $\left|\Delta p_x / p_\text{before}\right| $ \\
			$p_{x,\text{after}}$ & -9.196 & -7.410 & -16.606 & 1.86\% \\
			\hline $p_{y,\text{before}}$ & -2.192 & 0.083 & -2.108 & $\left|\Delta p_y / p_\text{before}\right| $ \\
			$p_{y,\text{after}}$ & -2.965 & 1.163 & -1.802 & 1.80\% \\
			\hline $L_{\text{before}}~(10^{-3}~
			\newton\,\meter\,\second)$ & 3.176 & 0.030 & 3.206 & $\left|\Delta L / L_\text{before}\right| $\\
			$L_{\text{after}}$ & 0.999 & 2.239 & 3.237 & 0.98\% 
		\end{tabular} 
		\label{tab:momentum} 
	\end{ruledtabular}
\end{table}

\begin{figure}
	\includegraphics{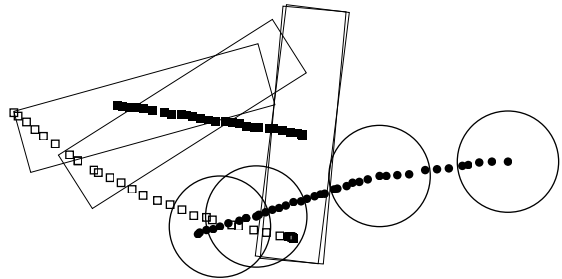}
\caption{Captured points with reconstructed object positions, showing the transfer of linear and angular momentum. The puck's initial velocity was 30~cm/s. Positions were sampled at 50~Hz.} 
\label{fig:strobe} 
\end{figure}

\section{Tracking Light Spots -- The Cavendish Experiment} 
\begin{figure}
	\includegraphics{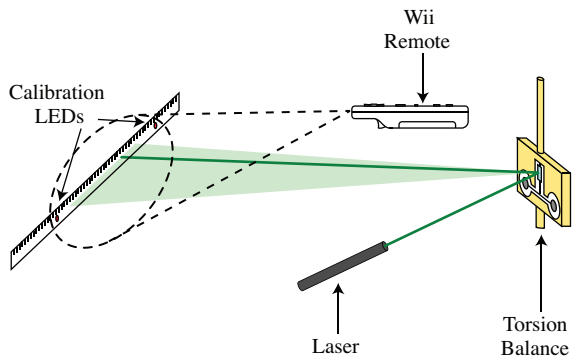} 
\caption{Experimental setup for the Cavendish experiment with Wiimote data acquisition.} 
\label{fig:cav-setup} 
\end{figure}
In the modern version of the Cavendish experiment, a laser is used to measure a torsion pendulum's rotation due to gravitational forces.\cite{Leybold} The typical procedure requires students to measure both the period of the oscillation and the change in the equilibrium position when the external masses are swiveled from one side of the apparatus to the other. Since the pendulum's period is on the order of several minutes, manually recording the position of the laser spot is tedious, especially if the data is to be taken over several runs.

A number of automated data acquisition techniques are available for this experiment.\cite{fischer:855,d'anci:348,fitch:2007} Few, however, are as simple or affordable as the Wiimote. The student mounts the Wiimote directed towards the reflected spot of the laser, and lets the computer record its position at regular intervals. As with the previous experiment, a reference is needed to translate Wiimote coordinates into actual lengths. As shown in Fig.~\ref{fig:cav-setup}, we have installed an LED on both sides of the laser's target. The Wiimote captures the laser spot and the LEDs simultaneously, and the software uses the known distance between the LEDs to calculate the actual distance traversed by the laser spot. Mounting the LEDs this way and remeasuring their position on every sample is a simple way to make the experiment self-calibrating. If the position of the Wiimote remains absolutely fixed, the appropriate scale factor could simply be hard coded. We effectively eliminate yaw by using a carpenter's square to set the side of the Wiimote housing perpendicular to the target wall.

\begin{figure}
	\includegraphics{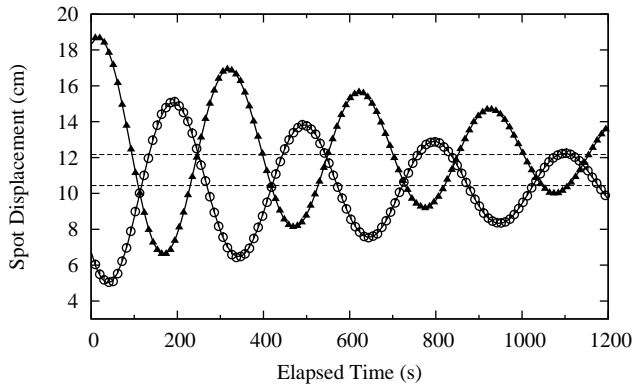} 
\caption{Student data obtained with the Wiimote-assisted Cavendish experiment. The dashed lines represent the central displacement obtained from the fits.} 
\label{fig:cav-fits} 
\end{figure}

Figure~\ref{fig:cav-fits} displays the data for two runs, clearly showing the expected damped harmonic motion. Each data set is obtained with the external lead masses in one of two possible orientations. The curves are fit to exponentially damped sinusoids. The period and equilibrium offset from this data set is consistent with a value of $G=\left(6.72 \pm 0.17\right)\times10^{-11}~
\newton\meter\cubed\per\kilogram\squared$. The measured value deviates from the standard value by only 0.77\%. It is important to note that the contribution to the total error from this data acquisition scheme is an order of magnitude smaller than the contribution from uncertainties in the geometry of the apparatus, such as the position of the lead ball. 

\section{Conclusion}

Our two examples suggest that the Wiimote can be a useful addition to many different kinds of experiments. The primary requirement is to make the intended light sources visible to the Wiimote while keeping distracting light sources out of its view. With the options we have presented (LEDs, lasers, and reflectors), there is likely a way to incorporate the Wiimote into most small-scale motion experiments. Larger experiments, such as those involving projectile motion, may face tougher challenges.

For appropriate experiments, the Wiimote is a versatile way to collect abundant, high-quality data. Its price is typically a fraction of dedicated, commercial data acquisition tools, making it ideal for use even in high schools. Moreover, its ease of use can enhance the appeal of dated or tedious experiments by inviting students to think about clever methods of data acquisition, and perhaps for the first time, to try their hands at computer programming.

The authors would like to thank students Alisha Kundert, Andrew Levin, Ridwan Rahman, and Cody Simons for their data on each experiment. Student support was provided by the Vanderbilt University College of Arts and Science.

\appendix 
\section{Technical Details \label{sec:app}} Our work was done on Ubuntu Linux using an \emph{Iogear} Bluetooth 2.1 USB dongle. The momentum experiment was conducted on a Ealing air table system. The colliding pucks were covered with black construction paper, and reflective stickers from REI.

The Cavendish experiment used a 5~mW green diode laser with an external power supply and integrated heat sink (model MGM3 from Beta Electronics). Laser pointers work fine for quick experiments, but our experience showed that they do not hold up to the hours of continuous operation required by the Cavendish experiment.

Our versions of the Wiimote interface libraries, both the original and the one modified for weak (reflective) sources, are available online.\cite{supp}


\begin{thebibliography}{10}
\bibitem{nintendo}
    Manufacturer's website: $<$\url{http://www.nintendo.com}$>$
\bibitem{Vannoni2007}
  Maurizio~Vannoni and Samuele~Straulino, ``Low-cost accelerometers for physics
  experiments,'' European J.~Phys. \textbf{28}, 781--787 (2007)
\bibitem{ochoa:16}
    Romulo~Ochoa, Frank~G.~Rooney, and William~J.~Somers, ``Using the Wiimote in introductory physics experiments,''
  Phys.~Teach. \textbf{49}, 16--18 (2011)
\bibitem{kawam:508}
    Alae Kawam and Minjoon Kouh, ``Wiimote experiments: 3-D inclined plane problem
  for reinforcing the vector concept,'' Phys.~Teach. \textbf{49}, 508--509 (2011)
\bibitem{gtk2008}
    GTK Wiimote Whiteboard, $<$\url{http://www.stepd.ca/gtkwhiteboard/}$>$
\bibitem{williamson:2000}
    J.~Charles Williamson, Ramon~O. Torres-Isea, and Craig~A. Kletzing, ``Analyzing linear and angular momentum conservation
  in digital videos of puck collisions,'' Am.~J.~Phys. \textbf{68}, 841--847 (2000)
\bibitem{Leybold}
    \textit{Leybold Physics Leaflets}, H\"{u}rth, Germany
\bibitem{fischer:855}
    C.~W. Fischer, J.~L. Hunt, and P.~Sawatzky, ``Automatic recording for the Cavendish balance,'' Am.~J.~Phys. \textbf{55}, 855--856 (1987)
\bibitem{d'anci:348}
    A.~M. D'Anci and C.~E. Armentrout, ``A light-beam data recorder for determination of
  the gravitational constant: Anomalous driven oscillations of a gravitation
  torsion balance,'' Am.~J.~Phys. \textbf{56}, 348--351 (1988)
\bibitem{fitch:2007}
    Noah Fitch,  Wesley Bliven, and Tyler Mitchell, ``Automating data acquisition for the Cavendish balance to improve the measurement of G,'' Am.~J.~Phys. \textbf{75}, 309--312 (2007)
    \bibitem{supp} See supplementary material at [URL will be inserted by AIP] for downloadable computer code.
\end{thebibliography}
\end{document}